\documentclass[12pt]{article}

\usepackage[margin=1in]{geometry}
\usepackage{setspace}
\usepackage{graphicx}
\usepackage{amsmath}
\usepackage{enumerate}
\usepackage{apacite}
\usepackage{color}
\usepackage{multirow}
\usepackage{url}
\usepackage[labelfont=bf]{caption}

\begin{document}
\pagenumbering{roman}
\title{Bayesian Probabilistic Projection of International Migration Rates}
\author{Jonathan J. Azose and Adrian E. Raftery\textsuperscript{1} \\
Department of Statistics \\
University of Washington}
\footnotetext[1]{Jonathan J. Azose is a Graduate Research Assistant and Adrian E. Raftery is a Professor of Statistics and Sociology, both at the Department of Statistics, Box 354322, University of Washington, Seattle, WA 98195-4322 (Email: jonazose@u.washington.edu/raftery@u.washington.edu). This work was supported by the Eunice Kennedy Shriver National Institute of Child Health and Development through grants nos. R01 HD054511 and R01 HD070936, and by a Science Foundation Ireland E.~T.~S.~Walton visitor award, grant reference 11/W.1/I2079. The authors are grateful to Patrick Gerland and Joel Cohen for sharing data and helpful discussions.}
\date{October 26, 2013}
\maketitle 

\newpage

\begin{abstract}
We propose a method for obtaining joint probabilistic projections of migration rates for all countries, broken down by age and sex. Joint trajectories for all countries are constrained to satisfy the requirement of zero global net migration. We evaluate our model using out-of-sample validation and compare point projections to the projected migration rates from a persistence model
similar to the method used in the United Nations' {\it World Population
Prospects}, and also to a state of the art gravity model. 
We also resolve an apparently paradoxical discrepancy between growth trends 
in the proportion of the world population migrating and the average 
absolute migration rate across countries. \\

\end{abstract}

\newpage
\baselineskip=18pt

\section{Introduction}\label{sec:intro}
\pagenumbering{arabic}

\begin{figure}
\centering
\includegraphics[width=1.0\textwidth]{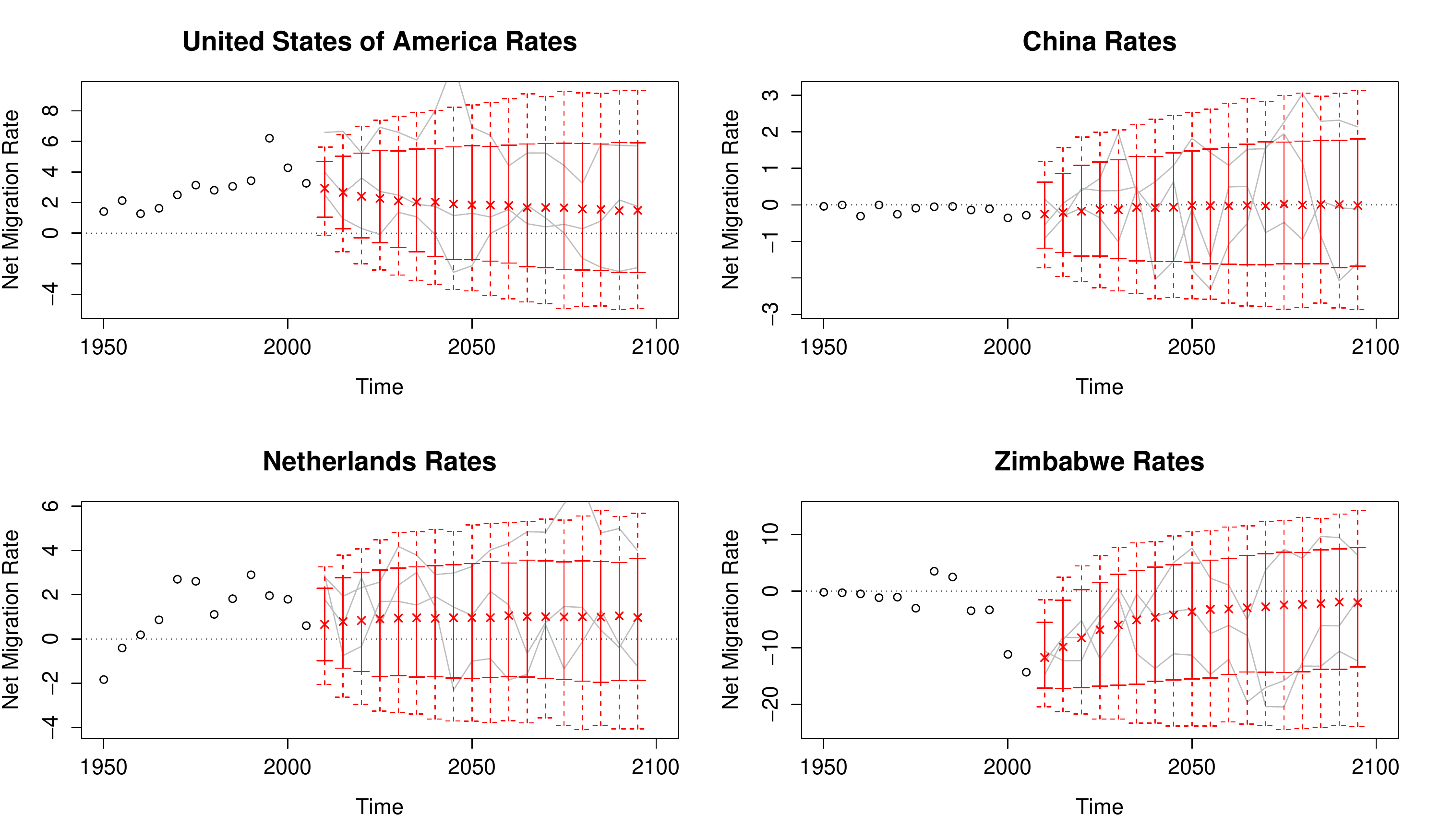}
\caption{Probabilistic Projections of Net International Migration Rates:
80\% and 95\% prediction intervals for four countries, with example trajectories included in gray.}
\label{fig:trajectories}
\end{figure}

In this paper we propose a method for probabilistic projection of net international migration rates. Our technique is a simple one that nonetheless overcomes some of the usual difficulties of migration projection. First, we produce both point and interval estimates, providing a natural quantification of uncertainty. Second, since our model uses only demographic variables as inputs, we can make long-term projections without explosion in the degree of uncertainty. Third, simulated trajectories from our model satisfy the common sense requirement that worldwide net migration sum to zero for each sex and age group. Fourth, our projected trajectories approximately replicate the observed frequency of countries switching between positive and negative net migration. Lastly, we sidestep the difficulty in projecting a complete large matrix of pairwise flows by instead working directly with net migration rates. Sample projections from our model for several countries are given in Fig. \ref{fig:trajectories}.

We also highlight an apparent paradox in the evolution of migration trends over time. We provide an explanation for this paradox and show that our model successfully reproduces it.

In the remainder of Section \ref{sec:intro}, we provide background and describe global trends in migration rates. In Section \ref{sec:methods}, we describe our data and methods for producing probabilistic projections. Section \ref{sec:results} summarizes our main results, including an evaluation of our model's performance and what our projections predict about future global migration trends. Finally, Section \ref{sec:discussion} contains evaluative discussion.
 
\subsection{Motivation and background}

There is a clear demand for migration projections. Organizations including the United Nations and the UK Office for National Statistics have identified a necessity for migration forecasts \cite{united2011,wright2010}. 

Our work is motivated by the needs of the UN Population Division in
producing probabilistic population projections for all countries.
The UN has recently adopted a Bayesian approach to projecting the
populations of all countries as the basis for its official medium projection,
and has issued probabilistic projections on an experimental basis
\shortcite{raftery2012population,united2013}, 
The underlying method can account
for uncertainty about fertility and life expectancy though
Bayesian hierarchical models \shortcite{alkema2011,raftery2012life}.
However, the approach does not yet take account of uncertainty
about international migration. 
Instead the UN probabilistic population projections are conditional
on deterministic migration projections that essentially amount to assuming that
current migration rates will continue into the future in the medium term.
To make the method fully probabilistic would require 
probabilistic projections of net international migration for all countries.

\citeA{lutz2004}, in answering the question of how to deal with uncertainty in population forecasting, point to the need for simple approaches to probabilistic forecasting of migration. Our paper attempts to meet this need. Despite the demand, some experts have been pessimistic about the possibility of predicting migration at all. \citeA{heide1963} felt that the task of finding a usable model for migration is ``virtually impossible''. This opinion was updated by \citeA{bijak2010}, who drew the similarly disheartening conclusions that ``migration is barely predictable'' and ``forecasts with too long horizons are useless.''

Nevertheless, there have been efforts to forecast international migration. These attempts have mostly been limited in geographic and/or chronological scope. \citeA{bijak2010} produced migration projections for seven European countries until 2025 using Bayesian hierarchical models. Using another geographically focused method, \citeA{fertig2000} projected migration flows from a set of 17 mostly European countries to Germany over the 1998-2017 time period. 
One drawback of these approaches in the context of population projections for {\em all} countries is that both 
require the use of data on migration flows between pairs of countries. Estimates of reasonable quality of these flows are now available for most pairs of European countries \cite{abel2010}, making such techniques feasible for Europe and other developed regions. Estimates for global pairwise migration flows are also available \cite{abel2013}, but the quality of these estimates varies with the reliability of record keeping in the countries involved. 

Another forecasting method was provided by \citeA{hyndman2008}, who gave a stochastic model for indirect migration forecasting by forecasting fertility and mortality, taking migration to be the appropriate quantity to satisfy the balancing equation. Their method provides estimates for individual countries, but joint estimates for all countries would in general not satisfy the requirement that worldwide net migration be zero. A simpler approach is taken by the United Nations World Population Prospects \citeyear{united2013}, which includes point projections that generally project migration rates to persist at or near current levels for the next couple of decades and drop deterministically to zero in the long horizon. Finally, \citeA{cohen2012} provides a method for point projections of migration counts for all countries using a gravity model.

\subsection{Theory of International Migration}

There is a general consensus about the major causes of international migration. On the individual level, desire to migrate is caused in large part by economic factors \shortcite{gallup2011,massey1993}. 
Refugee movements may be precipitated by political or social factors rather than economic ones \cite{richmond1988}. 
However, both economic and political factors are unlikely to be predictable in the long run with any useful degree of certainty. For the purposes of projection, \citeA{kim2010} argue for the use of more predictable demographic variables in place of unpredictable economic ones. They propose a model for prediction of migration flows which incorporates life expectancy, infant mortality rate, and potential support ratio as predictor variables. \citeauthor{kim2010} find these variables to be significant predictors of migration flows. 
Furthermore, as demographic variables tend to change much more slowly
than economic or political ones, it is often possible to project the values of demographic variables decades into the future with a lower degree of uncertainty. Our model projects net migration rates on the basis of only past migration rates and projected populations for all countries, 
for which forecasts can be made with enough precision to be useful.

One further demographic variable of interest in modeling migration is age structure. Age structure is important to migration modeling in two different ways. First, projected age structures for all countries can potentially be used as predictor variables in projections of future migration. Since labor migration is common, the age structure of the sending and/or receiving countries can be used in making projections \cite{fertig2000,hatton2002,hatton2005}.
\citeA{kim2010}, in a study of pairwise migration flows, found that a young age structure in the country of origin is associated with high migration flows, while a young age structure in the country of destination is associated with low flows. 

Second, it may be of interest to project not only net migration rates, but also \emph{age-specific} net migration rates. 
\citeA{rogers1981} provided a parametric multiexponential model migration schedule which can be used in converting from projected net migration rates to age-specific rates. Their model incorporates a principal migration peak among young adults, who often migrate for reasons of economics, marriage, or education, as well as a secondary childhood peak for the children of those young adult migrants. They include a further option for waves of retirement and post-retirement migration which are common patterns of regional migration but less common internationally. \citeA{raymer2007} point out the complication that the age structure of a migrating population is dependent on direction of migration. For example, we would expect a labor migration and a subsequent return migration to have different age structures. This fact is unfortunately difficult to incorporate into a model like ours which works with net rates rather than gross pairwise flows.

For projection purposes, Bayesian modeling is well suited to modeling international migration. The difficulty in making accurate point projections emphasizes the need for an approach that produces estimates of uncertainty. As our data set includes only 12 time points per country, non-Bayesian inference could be difficult; the Bayesian approach alleviates this by allowing us to borrow strength across countries. Studies with limited geographical scope confirm this intuition. In a comparison of several methods for forecasting migration to Germany, \citeA{brucker2006} found performance of a hierarchical Bayes estimator to be superior to that of a simpler OLS estimator. Good results have also come out of Bayesian forecasting efforts for fertility and mortality \cite{alkema2011,LalicRaftery2012,raftery2012population,raftery2012life}. In addition to forecasting, estimation of demographic variables also lends itself to Bayesian methodology \shortcite{abel2010,congdon2010,Wheldon&2013}.

\subsection{Migration trends}\label{sec:trends}

\begin{figure}
\centering
\includegraphics[width=1.0\textwidth]{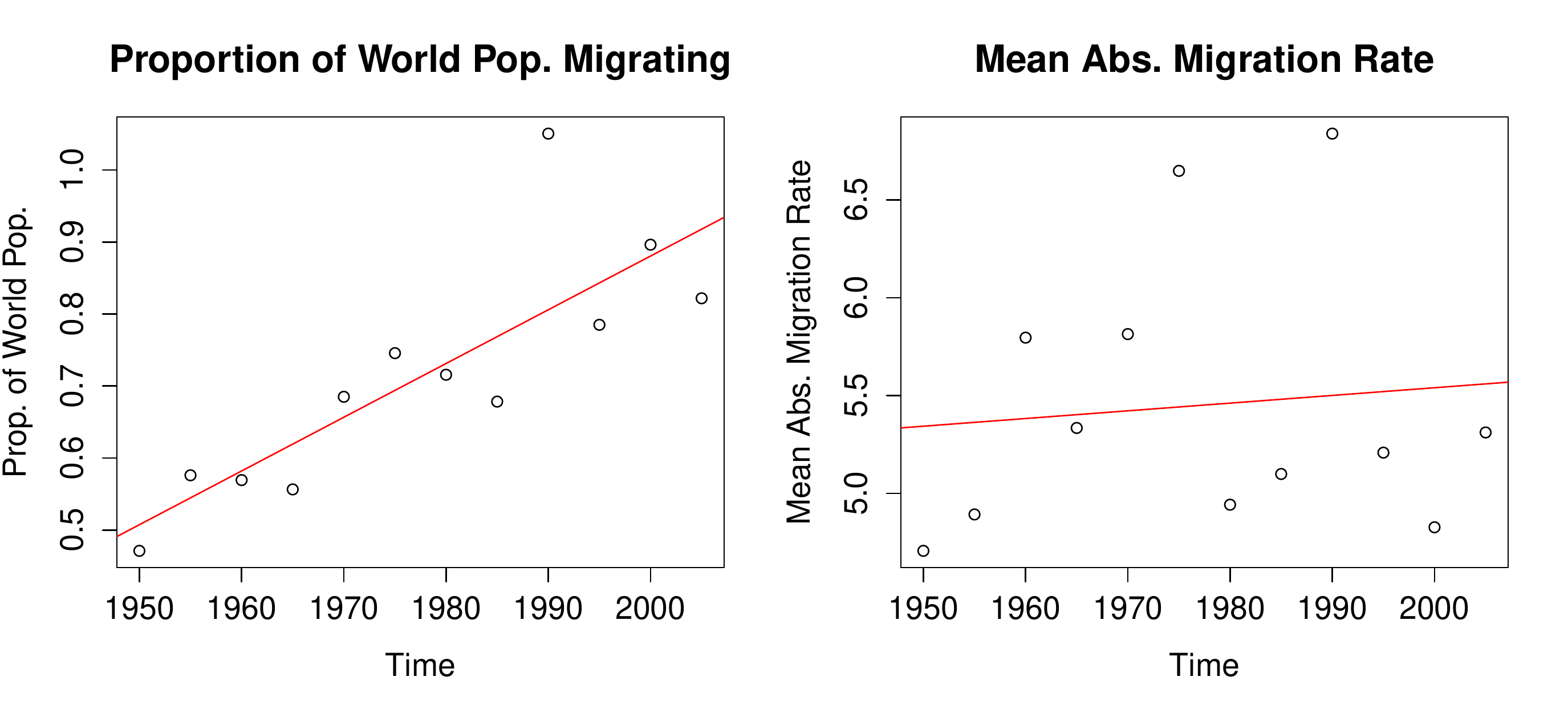}
\caption{Global Trends in International Migration:
Left: Time series of the estimated proportion of the world population migrating.
Right: Average absolute migration rate, averaged across all countries
at each time point.
Both plots show number of migrants per thousand population. 
The red lines are ordinary least squares regression lines.}
\label{fig:propMAMR3}
\end{figure}

The primary goal of our model is to produce point and interval projections. However, it is also desirable for our model to replicate current trends in the migration data. When looking at migration trends over roughly the last 60 years, we find an apparent contradiction. Consider the question of whether migration increased between 1950 and 2010. One sensible way to answer this question is to look at the number of individuals migrating within each five-year time period per thousand individuals of the world population. We will denote this quantity by $prop(t)$.\footnote{To calculate this quantity, we used data that take the form of \emph{net} numbers of migrants per country rather than \emph{gross} counts. We made the approximation that most countries are either purely senders or purely receivers, so that gross numbers can be approximated by net numbers. For our purposes, what is important is that this approximation not become much better or worse over time.} The left panel of Fig. \ref{fig:propMAMR3} shows the trend in $prop(t)$ over the period from 1950 to 2010. There is a clear upward trend, with 74\% growth in $prop(t)$ between the 1950 time period and the 2005 time period. This growth is significant. A $t$-test shows strong evidence of non-zero slope ($p=0.00087$, $R^2=0.69$).

On the other hand, we might answer the question of whether migration is increasing over time at the country level rather than the global level. We can do so by computing the mean absolute migration rate, $mamr(t)$, averaged across all countries. The right panel of Fig. \ref{fig:propMAMR3} shows this trend over the period from 1950 to 2010. Whereas there was clear growth in $prop(t)$ over this time period, $mamr(t)$ shows a much smaller amount of growth, with only 13\% growth between the 1950 time period and the 2005 time period. A $t$-test does not show evidence of non-zero slope ($p=0.74$, $R^2=0.00005$).

Thus, there is an apparent contradiction: How is it possible that more people are migrating than in the past but countries' migration rates are not 
increasing on average? 
In Section \ref{sec:paradoxResolution} we resolve this paradox.

A second feature of the historical migration data to consider is the frequency with which countries switch between being net senders and net receivers of migrant. Such switches have been relatively common over the past 50 years. In fact, in the 2005-2010 time period, 46\% of countries had different migration parity than they had in 1955-1960 (i.e., they switched either from net senders to net receivers or vice versa.) In contrast, the current United Nations methodology \cite{united2013} projects \emph{no} crossovers between now and 2100. Our model projects crossover behavior that is more in line with historical trends. Further analysis of projected parity changes is given in Section \ref{sec:denmark}.

\section{Methods}\label{sec:methods}

\subsection{Data}

We use data from the 2010 revision of the United Nations Population Division's biennial \emph{World Population Prospects} (WPP) report \cite{united2011}. 
WPP reports contain estimates of countries' past age- and sex-specific 
fertility, mortality and net international migration rates, 
as well as projections of future rates.

The quantity we are interested in forecasting is $r_{c,t}$, the net annual migration rate for country $c$ in time period $t$, reported in units of migrants per thousand individuals in the WPP data. For calculations, we sometimes convert \emph{rates} $r_{c,t}$ to corresponding \emph{counts} $y_{c,t}$. Our method also requires knowledge of the average population of countries, $n_{c,t}$, indexed by country and time, and projections of $n_{c,t}$ into the future for all countries.

\subsection{Probabilistic Projection Method}

Our technique is to fit a Bayesian hierarchical first-order autoregressive, or AR(1), model to net migration rate data for all countries. We model the migration rate, $r_{c,t}$, in country $c$ and time period $t$ as
$$(r_{c,t}-\mu_c)=\phi_c(r_{c,t-1}-\mu_c)+\varepsilon_{c,t},$$
where $\varepsilon_{c,t}$ is a normally-distributed random deviation 
with mean zero and variance $\sigma^2_c$. We put normal priors 
on each country's theoretical equilibrium migration rate $\mu_c$, and a uniform prior on the autoregressive parameter $\phi_c$. 
Under this model, simulation of trajectories requires us to estimate or specify values of $\mu_c, \phi_c$, and $\sigma^2_c$ for all countries, so the complete parameter vector is given by $\boldsymbol{\theta}=(\mu_1, \ldots, \mu_C, \phi_1, \ldots, \phi_C, \sigma^2_1, \ldots, \sigma^2_C)$, where $C$ is the number
of countries.

The full specification of the model, including prior distributions,
is as follows:

$$
\textrm{Level 1}
\left\{
\begin{array}{l}
(r_{c,t}-\mu_c)=\phi_c(r_{c,t-1}-\mu_c)+\varepsilon_{c,t}\\
\varepsilon_{c,t} \overset{\textrm{ind}}{\sim} N(0,\sigma^2_c)
\end{array}
\right.
$$

$$
\textrm{Level 2}
\left\{
\begin{array}{l}
\phi_c \overset{\textrm{iid}}{\sim} U(0,1)\\
\mu_c \overset{\textrm{iid}}{\sim} N(\lambda, \tau^2)\\
\sigma^2_c \overset{\textrm{iid}}{\sim} IG(a,b)\\
\end{array}
\right.
$$

$$
\textrm{Level 3}
\left\{
\begin{array}{l}
a \sim U(1,10)\\
b|a \sim U(0,100(a-1))\\
\lambda \sim U(-100,100)\\
\tau \sim U(0,100),
\end{array}
\right.
$$
where $X \sim N(\mu,\sigma^2)$ indicates that the random variable 
$X$ has a normal distribution with mean $\mu$
and variance $\sigma^2$ (and hence standard deviation $\sigma$),
$U(c,d)$ denotes a uniform distribution between the limits $c$ and $d$,
and $IG(a,b)$ denotes an inverse gamma distribution with 
probability density function (as a function of $x$) proportional to
$x^{-a-1} e^{-b/x}$.

We obtain draws from the posterior distributions of all parameters using Markov Chain Monte Carlo methods. In our implementation, we use the Just Another Gibbs Sampler (JAGS) software package for Markov chain Monte Carlo simulations 
\cite{plummer2003}.

Having obtained a sample $(\boldsymbol{\theta}_1, \ldots, \boldsymbol{\theta}_N)$
of draws from the joint distribution of the parameters, we use these draws to obtain a sample from the joint posterior predictive distribution. For each sampled point $\boldsymbol{\theta}_k$ from the joint posterior distribution of the parameters, we first simulate a set of joint trajectories $\tilde r_{c,t}^{(k)}$ for net migration rates at time points until 2100,
where $k$ indexes the trajectory.
However, this procedure generally produces trajectories which are impossible in that they give nonzero global net migration counts. We therefore create corrected net migration rate trajectories $\tilde r_{c,t}^{*(k)}$, using the 
following method:

\begin{enumerate}[1.]
\item On the basis of the parameter vector $\boldsymbol{\theta}_k$, project net migration rates for all countries a single time point into the future. Denoting the next time period in the future by $t'$, this allows us to obtain a collection of (uncorrected) projected values $\tilde r_{c,t'}^{(k)}$ for all countries $c$.
\item Convert net migration rate projections $\tilde r_{c,t'}^{(k)}$ to net migration count projections $\tilde y_{c,t'}^{(k)}$. This is done by multiplying by a projection of each country's population, $\tilde n_{c,t'}$. We obtain these projections from WPP 2010 \cite{united2011}.
\item Further break down migration counts by age $a$ and sex $s$ to obtain estimates of net male and female migration counts for all countries and age groups, $\tilde y_{c,t',a,s}^{(k)}$. This is done by applying projected model migration schedules to all countries. 
We take each country's age- and sex-specific migration schedule to be the same as the distribution of migration by age and sex in the most recent time point for which detailed data were available for that country. 
\item For each simulated trajectory, within each age and sex category, apply a correction to ensure zero worldwide net migration. The correction we apply redistributes any overflow migrants to all countries, in proportion to their projected populations. Specifically, take the corrected migration count projection $\tilde y_{c,t',a,s}^{*(k)}$ to be
$$\tilde y_{c,t',a,s}^{*(k)}=\tilde y_{c,t',a,s}^{(k)}-\frac{\tilde n_{c,t'}}{\sum_{j=1}^C \tilde n_{j,t'}} \sum_{j=1}^C \tilde y_{j,t',a,s}^{(k)}.$$
\item Convert the corrected age- and sex-specific net migration counts $\tilde y_{c,t',a,s}^{*(k)}$ back to corrected net migration rates $\tilde r_{c,t'}^{*(k)}$ by disaggregating and converting counts to rates.
\item Continue projecting trajectories one time step at a time into the future by repeating steps 1-5.
\end{enumerate}

Note that, although the uncorrected net migration rates $\tilde r_{c,t'}$ come from the desired marginal posterior predictive distributions, the correction in step 4 changes those distributions by projecting them onto a lower dimensional space. Sensitivity analysis suggests that the correction introduces only minor changes between the marginal distributions with and without the correction.

\section{Results}\label{sec:results}

\subsection{Evaluation}

We do not know of any other model that produces \emph{probabilistic} projections of all countries' migration rates. However, we can take our model's median projections to be point projections and compare them with models that produce point projections only. First, as a baseline for comparison, we evaluate them against the simple \emph{persistence model} which projects migration rates to continue at the most recently observed levels indefinitely into the future. In the short to medium horizon, the persistence model is similar to the expert knowledge-based projections in the WPP \cite{united2011}.

Second, we compare against point projections produced separately for all countries using the gravity model based method of \citeA{cohen2012}. The gravity model produces projected migration counts, but we convert these to rates for comparability with our method. For each country $c$, the gravity model makes projections as follows: Let $L(t)$ be the population of country $c$ at time $t$, and let $M(t)$ be the population of the rest of the world at time $t$. Then expected in-migration to country $c$ is given by $a \times L(t)^\alpha M(t)^\beta$, where $a$ is a country-specific proportionality constant. The exponents $\alpha$ and $\beta$ are constant across countries, with values estimated by \citeA{kim2010}. Similary, expected out-migration from country $c$ has the form $b\times L(t)^\gamma M(t)^\delta$, where $b$ is to be estimated and $\gamma$ and $\delta$ come from \citeA{kim2010}. The constants of proportionality $a$ and $b$ for each country are chosen to minimize the sum of squared deviations between estimates of net migration produced by the gravity model and true historical values of net migration from the WPP 2010 revision \cite{united2011}. Having estimated $a$ and $b$ for a particular country, net migration projections are then given by $a \times L(t)^\alpha M(t)^\beta - b\times L(t)^\gamma M(t)^\delta$, where $L(t)$ and $M(t)$ are now projected populations. Implementation details are given in the appendix.

Our historical data consist of a series of migration rates $r_{c,t}$ for 197 countries at 12 time points in five-year time intervals, spanning the period from 1950 to 2010. We performed an out-of-sample evaluation by holding out the data from the $m$ most recent time points for all countries and producing posterior predictive distributions on the basis of the remaining $(12-m)$ time points. As point forecasts we used the median of the posterior predictive distribution.
We report out-of-sample mean absolute error as a measure of the quality of point forecasts, and interval coverage as a measure of quality of our interval predictions.

Table \ref{tab:eval} contains these evaluation metrics for our Bayesian hierarchical model and the mean absolute errors for the persistence and gravity models.
Across the board, our point projections outperformed both the persistence model and the gravity model, and our interval projections achieved reasonably good calibration.

\begin{table}[h!]
\caption{Predictive Performance of Different Methods: Mean absolute errors (MAE) and prediction interval coverage for our Bayesian hierarchical model, the gravity model, and the persistence model.}\label{tab:eval}
\begin{center}
\begin{tabular}{|l|c|c|c|c|c|c|}
\hline
Validation time period & Model & MAE & 80\% Cov. & 95\% Cov.\\
\hline
\multirow{3}{*}{5 years} & Bayesian & 3.24 & 91.4\% & 96.4\%\\ 
& Gravity & 4.70 & --- & --- \\
& Persistence & 3.57 & --- & --- \\
\hline
\multirow{3}{*}{15 years} & Bayesian & 4.76 &  84.9\% & 93.4\% \\
& Gravity & 6.57 & --- & --- \\
& Persistence & 6.74 & --- & --- \\
\hline
\multirow{3}{*}{30 years} & Bayesian & 5.12 & 77.2\% & 89.3\%\\
& Gravity & 12.32 & --- & --- \\
& Persistence & 7.17 & --- & --- \\
\hline
\end{tabular}
\end{center}
\end{table}

\subsection{Paradox Resolution}\label{sec:paradoxResolution}

In this section, we resolve the apparent paradox that migration rates have been roughly constant when averaged across countries despite growing numbers of global migrants over time. We first provide an algebraic explanation for how the proportion of the world population migrating, $prop(t)$, can grow over time while the mean absolute migration rate, $mamr(t)$, stays roughly constant. 
We then check that this algebraic explanation is consistent with the observed data.

We are interested in the change in two numbers over time: the mean absolute migration rate,
$$mamr(t)=\frac{\sum_{c=1}^C |r_{c,t}|}{C},$$
and the proportion of the world's population migrating, defined here as
\begin{equation*}\label{eqn:prop}
prop(t)\approx \frac{1}{2} \frac{\sum_{c=1}^C |y_{c,t}|}{\sum_{j=1}^C n_{j,t}}
= \frac{1}{2} \sum_{c=1}^C |r_{c,t}| \frac{n_{c,t}}{\sum_{j=1}^C n_{j,t}}
= \frac{1}{2} \sum_{c=1}^C |r_{c,t}| \psi_{c,t},
\end{equation*}
where $\psi_{c,t}=\frac{n_{c,t}}{\sum_{j=1}^C n_{j,t}}$ is the proportion of the world population residing in country $c$ in time period $t$. (The factor of 1/2 is so that migrants are not double-counted as both immigrants and emigrants.) Thus, $mamr(t)$ and $prop(t)$ are both weighted averages of absolute migration rates. The former uses uniform weights across all countries and the latter weights countries proportionally to their size. 

The question of interest is how $prop(t)$ can experience steady growth and increase by 74\% between 1950 and 2010 while $mamr(t)$ oscillates and grows by only 13\%. From a purely algebraic perspective, there is no inherent contradiction in these two different weighted averages growing at different rates, so long as some combination of the following two things is true: (1) the weights $\psi_{c,t}$ are changing over time in such a way that growth in $\psi_{c,t}$ happens disproportionately among countries with high values of $|r_{c,t}|$, and (2) growth in absolute migration rate is somehow related to country size.

\begin{figure}
\centering
\includegraphics[width=\textwidth]{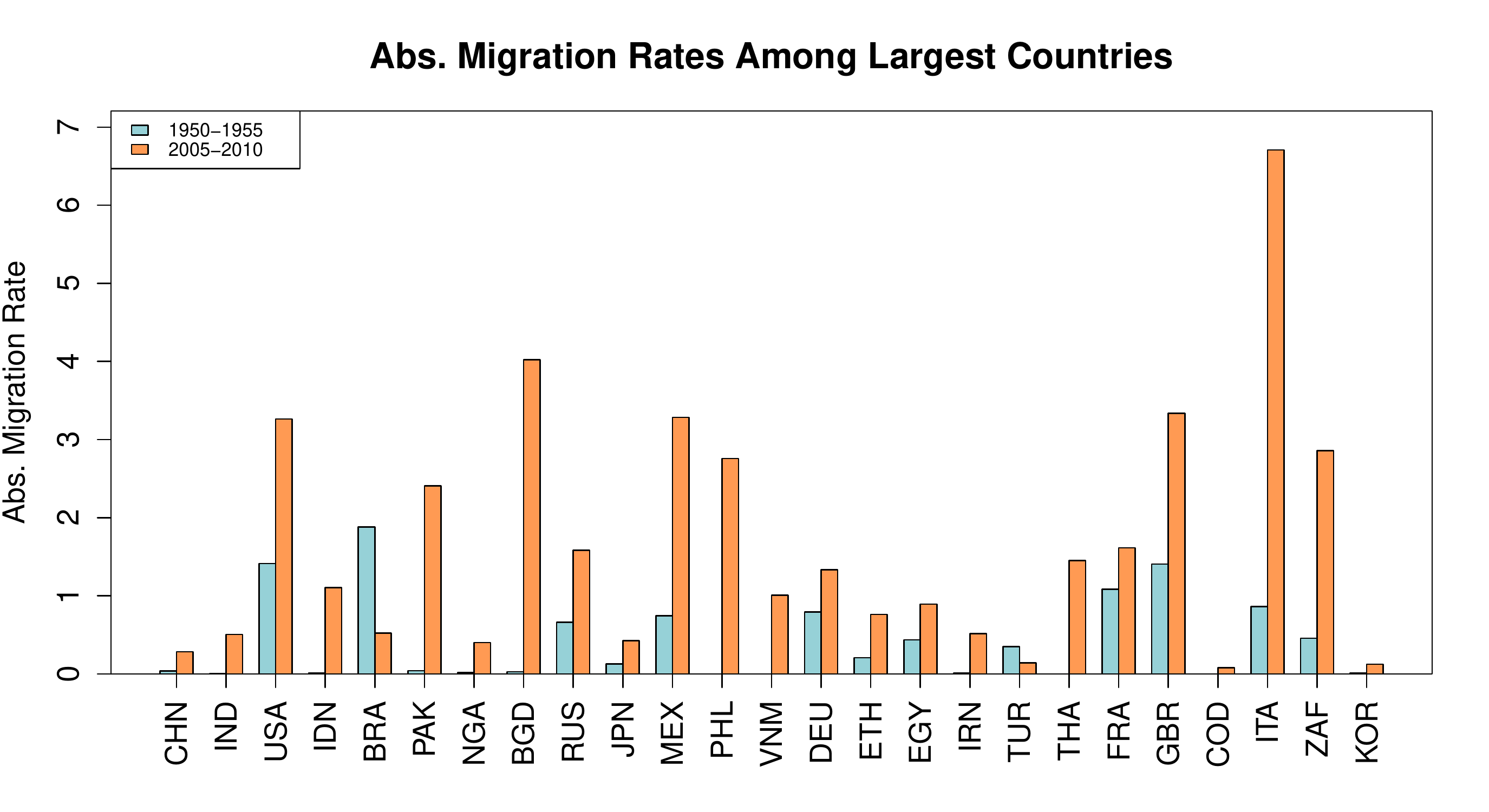}
\caption{Absolute annual migration rates per thousand individuals in the 25 most populous countries. Labels on the $x$-axis are three-letter ISO country codes.}
\label{fig:paradoxResolution}
\end{figure}

In fact the population-based weights $\psi_{c,t}$ do not change much over the period we are investigating. Countries which were large half a century ago are generally still large today. In fact, the growth in $prop(t)$ is mostly driven by growth in $|r_{c,t}|$ for the most heavily weighted (i.e. most populous) countries. Over the time period from 1950 to 2010, we see nearly across-the-board increases in absolute migration rates among the very highly populated countries. Figure \ref{fig:paradoxResolution} shows this growth among the largest countries. Orange bars show absolute migration rates for the 25 largest countries in 2005-2010, ordered from largest to smallest population. Blue bars show absolute migration rates from 1950-1955.

Of the 25 countries with the largest populations in 2005-2010, 23 had higher absolute migration rates in 2005-2010 than they did in 1950-1955. This collection of countries covers a majority of the world population---76\% of the world population in 1950-1955 and 75\% in 2005-2010. The mean absolute migration rate among the 25 largest countries was extraordinarily low in 1950-1955---only 0.42 per thousand, compared to a global average of 4.71 per thousand. By 2005-2010, the mean absolute migration rate among the 25 largest countries had grown to 1.74 per thousand against a global average of 5.31 per thousand. Notably, the mean absolute migration rate among large countries is still much lower than the worldwide average. Nevertheless, this small growth in absolute migration rates for the 25 largest  countries provides the majority of the increase in $prop(t)$. 

\begin{figure}
\centering
\includegraphics[width=1.0\textwidth]{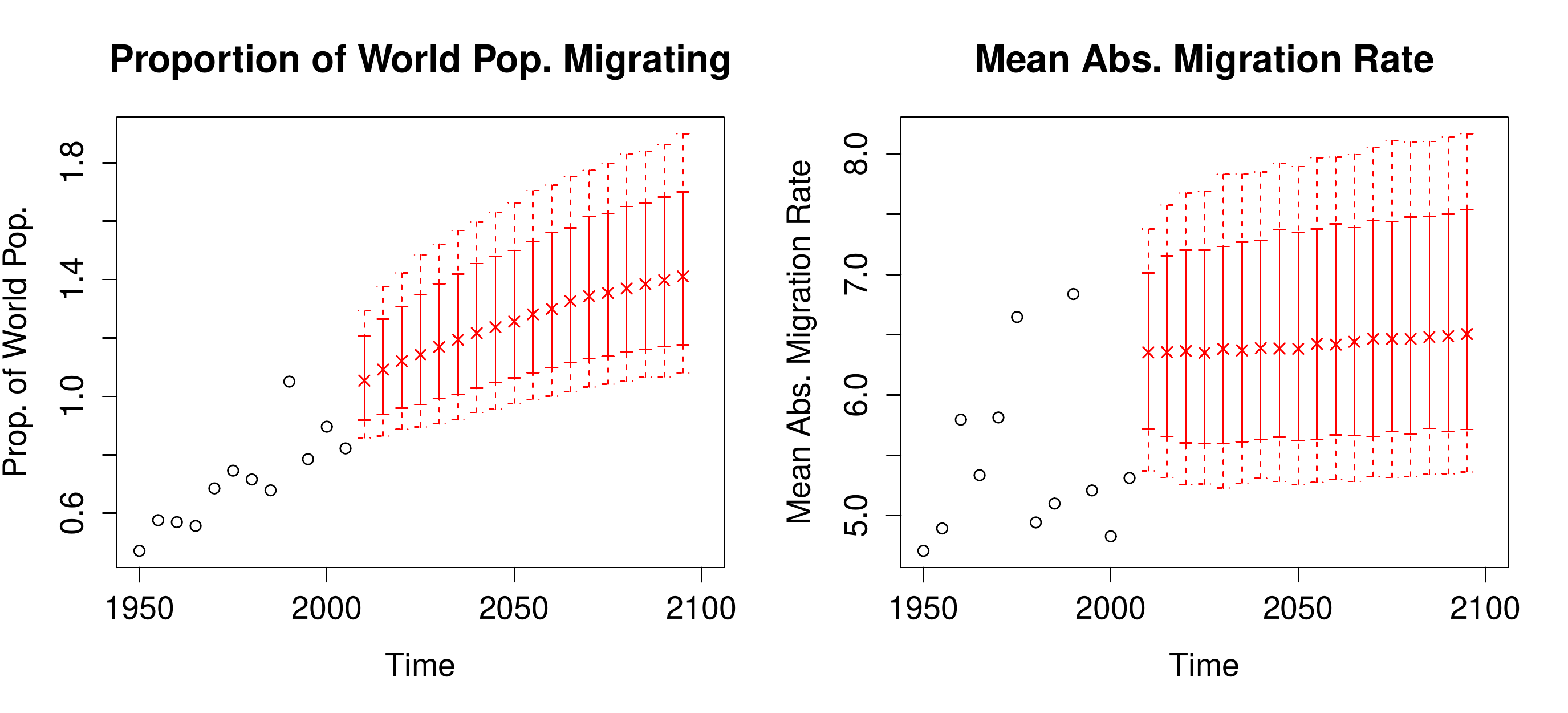}
\caption{In black, observed historical data on mean annual proportion of the world population migrating (left; per thousand) and mean absolute annual migration rate (right; per thousand) for five-year time periods from 1950 to 2010. In red, median estimates and 80\% and 95\% prediction intervals from our model for time periods out to 2100.}
\label{fig:propmamr2}
\end{figure}

The model we presented in Section \ref{sec:methods} produces projections that are consistent with the observed trends in $prop(t)$ and $mamr(t)$, despite containing no assumptions about or parameters directly tied to either $prop(t)$ or $mamr(t)$. Projections are shown in Fig. \ref{fig:propmamr2}. We forecast that $prop(t)$ will continue to grow, leveling off in the long horizon and that $mamr(t)$ will remain roughly constant. 

One way to interpret this projection is as a continued trend towards globalization. A defining feature of globalization is an increase in transnationalism in general which manifests itself by increases in cross-border flows of various kinds \cite{castles2003}. The continued growth of proportion of the world population migrating is therefore consistent with an increase in globalization. One result of globalization's characteristic transnational flows is an increase in homogeneity across nations \cite{robertson1992}. In this sense, too, our projections are consistent with an increase in globalization. Our model is projecting that net international migration rates among high-population countries will continue to converge towards those of the rest of the world.

\subsection{Case Studies}

\subsubsection{Denmark}\label{sec:denmark}

\begin{figure}
\centering
\includegraphics[width=0.6\textwidth]{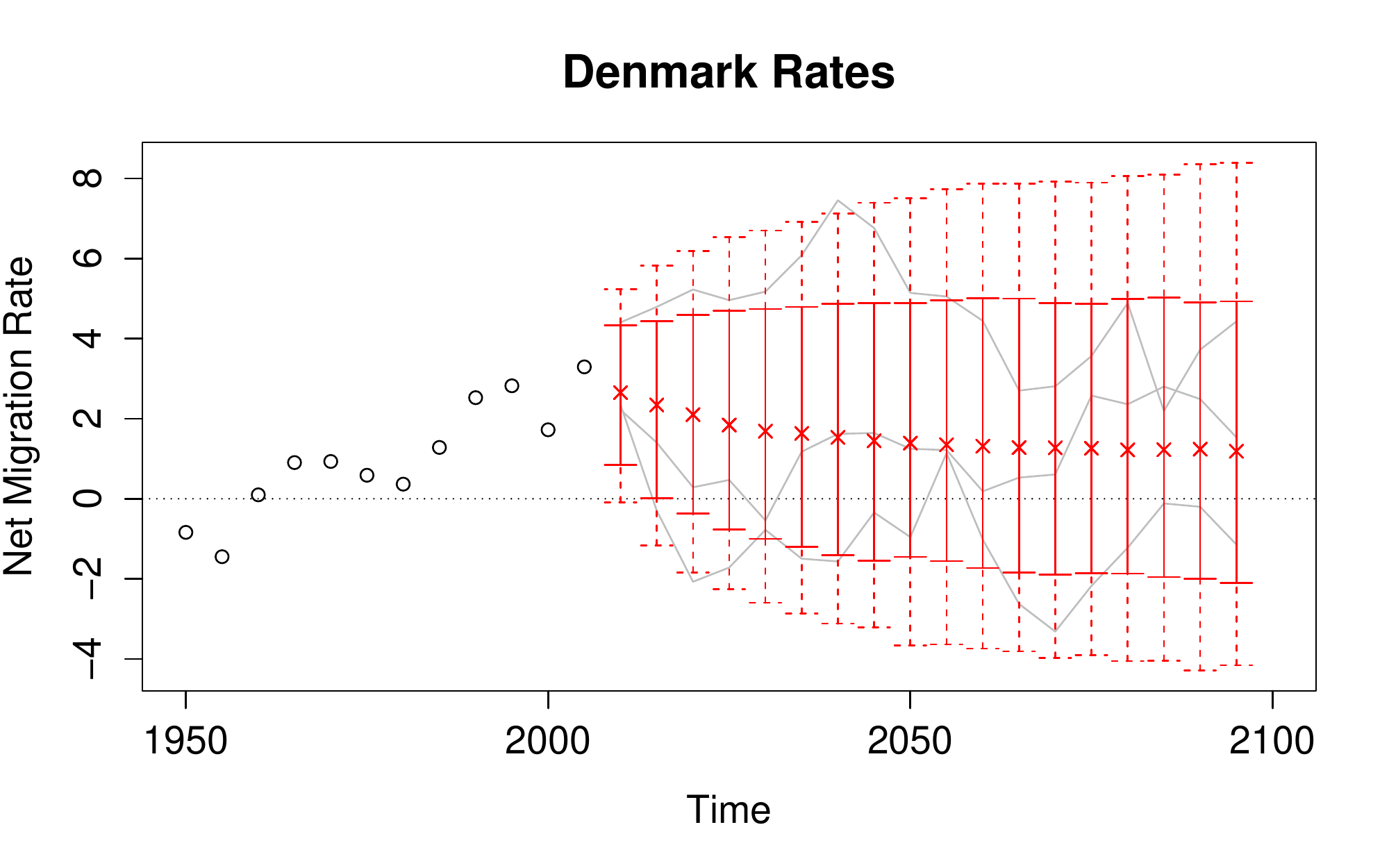}
\caption{Probabilistic Projections of Net International Migration Rates:
80\% and 95\% prediction intervals for Denmark, with example trajectories included in gray.}
\label{fig:denmark}
\end{figure}

Denmark experienced net emigration through the 1950s, but has consistently received net immigration since the 1960s. This pattern of changing from a net sender to a net receiver within the last 60 years is common to many of the European countries, including Norway, Finland, the UK, and Spain, among others. This serves as a reminder that the global migration to northern and western Europe which seems so firmly established now is a relatively recent phenomenon.

Our median predictions for Denmark have the country continuing to be a net receiver of migrants for as far out into the future as we care to project. However, we also see that the probability of Denmark switching over to a net sender increases over time. Based on the history of the 20th century, it seems realistic to include the possibility of changeovers in Denmark and other European countries in probabilistic migration projections. Correspondingly, projections that do not take account of this possibility seem unrealistic.

The European countries are not alone in having oscillated between being net senders and net receivers of migrants. As mentioned in Section \ref{sec:trends}, 46\% of countries had different migration parity in the 1955-1960 time period than they had in 2005-2010 (i.e., they switched either from net senders to net receivers or vice versa.) Our Bayesian hierarchical model projects 49\% of countries will have different migration parity in 2055-2060 than they do now. This projection is in line with the number of historical parity changes. In contrast, the gravity model \cite{cohen2012} projects only 29\% of countries to change parity by 2055-2060. The persistence model and the WPP migration projections \cite{united2013} both project no parity changes.

\subsubsection{Nicaragua}

Migration rates in Nicaragua have increased steadily in magnitude over the last six decades. Nevertheless, although our model projects a small probability of continued growth in the magnitude of the net migration rate, it gives higher probability to scenarios in which migration rates move back towards zero. In general, our model favors trajectories in which net migration rates move towards zero rather than continuing current trends of growth in magnitude where such trends exist. 

\begin{figure}
\centering
\includegraphics[width=0.6\textwidth]{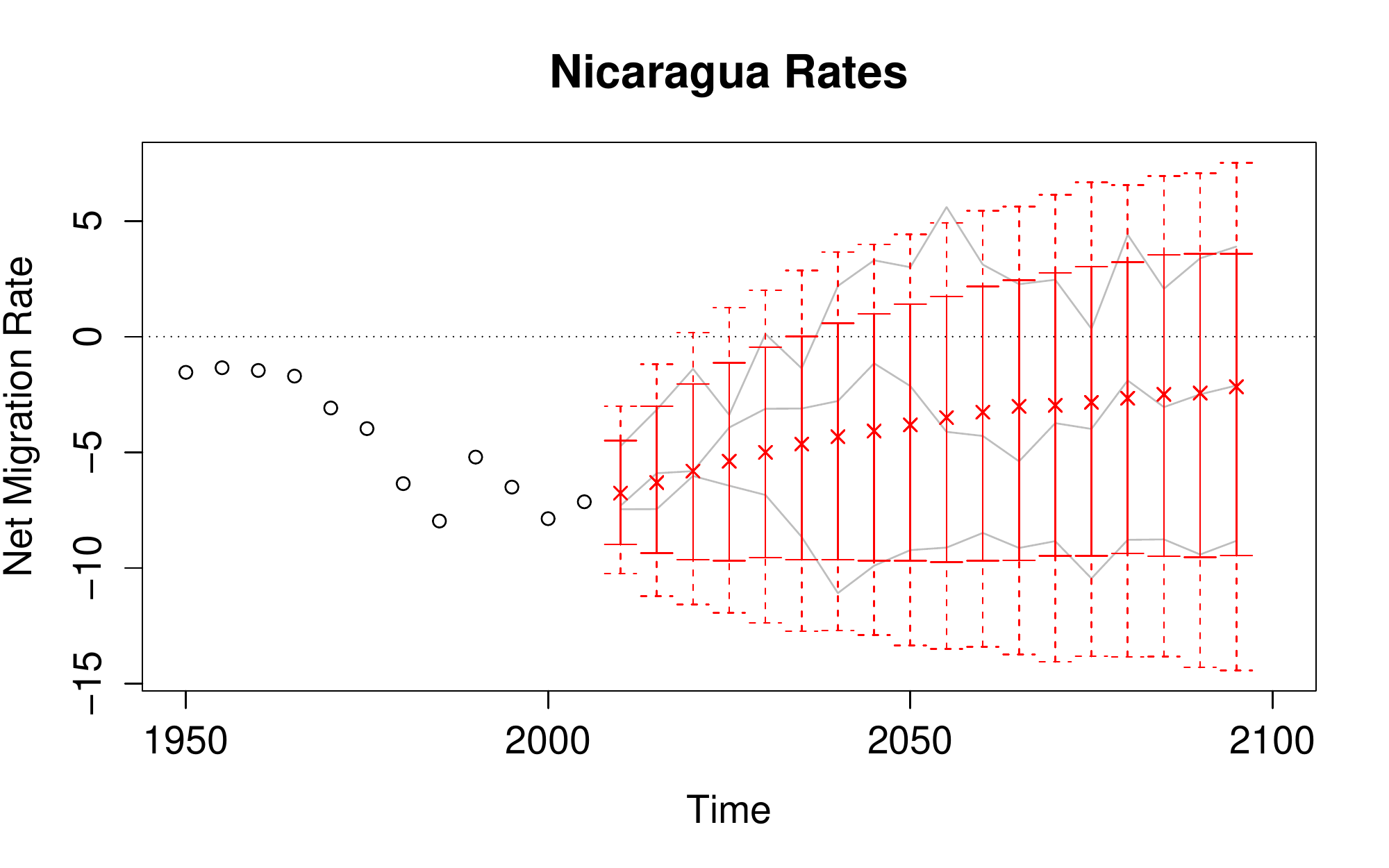}
\caption{Probabilistic Projections of Net International Migration Rates:
80\% and 95\% prediction intervals for Nicaragua, with example trajectories included in gray.}
\label{fig:nicaragua}
\end{figure}

Statistically, this tendency for migration rates on average to reverse course and tend back towards zero arises from the hierarchical nature of the model. Specifically, all of the $\mu_c$ values, which we can think of as the long-horizon median migration rates for each country, are assumed to come from a common $N(\lambda,\tau^2)$ distribution. As a result, the hierarchical ``sharing of strength'' has a tendency to pull all the $\mu_c$ values towards a common center, $\lambda$, which has a posterior distribution with a mode close to zero.
It should be noted that while our model's median projections tend to predict reversal in growth trends, the predictive probability distributions give substantial probability to continuation and growth of rates.

\subsubsection{India}

Historically, India has had relatively small net migration rates, on the order of less than 1 per thousand. The 95\% prediction intervals from our model are quite a bit wider than the range of India's historical data, expanding out to roughly $\pm 3$ per thousand.

\begin{figure}
\centering
\includegraphics[width=0.6\textwidth]{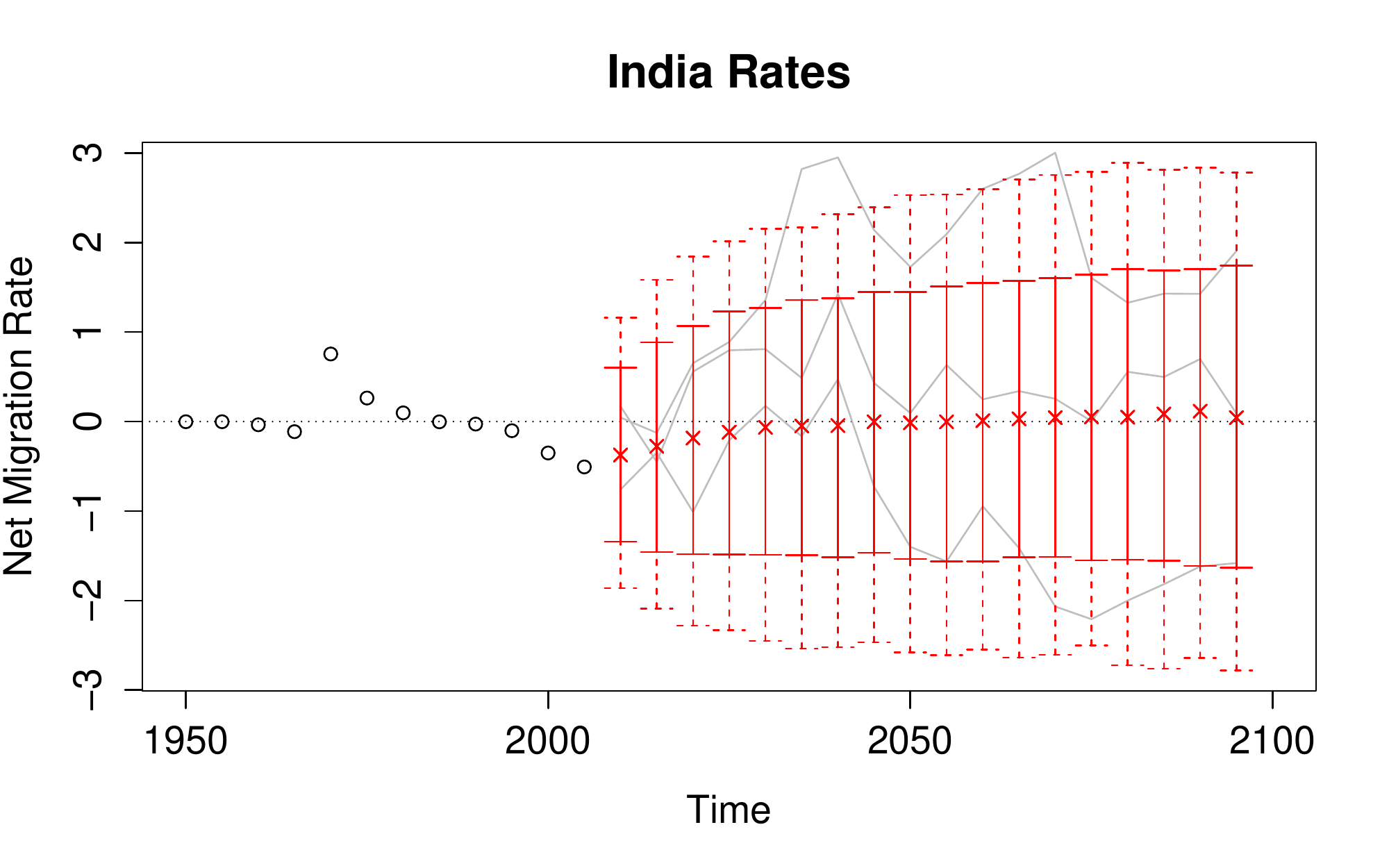}
\caption{Probabilistic Projections of Net International Migration Rates:
80\% and 95\% prediction intervals for India, with example trajectories included in gray.}
\label{fig:india}
\end{figure}

Statistically, the width of a country's prediction intervals from our model is primarily controlled by the error variance $\sigma^2_c$. (The autoregressive parameters, $\phi_c$, also influence the width of prediction intervals, but to a lesser extent.) The excess width of India's prediction intervals above its range of observed migration history is statistically a result of the hierarchical ``sharing of strength''. Since most other countries have larger ranges of migration rates, India's posterior distribution on $\sigma^2_c$ gets inflated somewhat to values more in line with the rest of the world. The same inflation of $\sigma^2_c$ occurs in China, which also has experienced uncommonly small migration rates in the past. 

Substantively, this seems realistic given the increasing globalisation we have documented. As the largest countries become more like other countries in terms of migration patterns, it seems reasonable to expect that the variability of their migration rates in the future would also increase to become more like the levels of other countries.

\subsubsection{Rwanda}

In the early 1990s, Rwanda experienced high net out-migration, followed by high net in-migration in the late 1990s. These migration spikes were a result of emigration during the Rwandan genocide in 1994 and subsequent return migration. Outside of the 1990s, Rwanda had quite small and stable migration rates. This pattern of stability punctuated by large shocks poses a problem for probabilistic projections: Do we get better performance with wide prediction intervals which encompass the high migration rates during the shock, or narrow prediction intervals which reflect the decades of stability around it?

\begin{figure}
\centering
\includegraphics[width=0.6\textwidth]{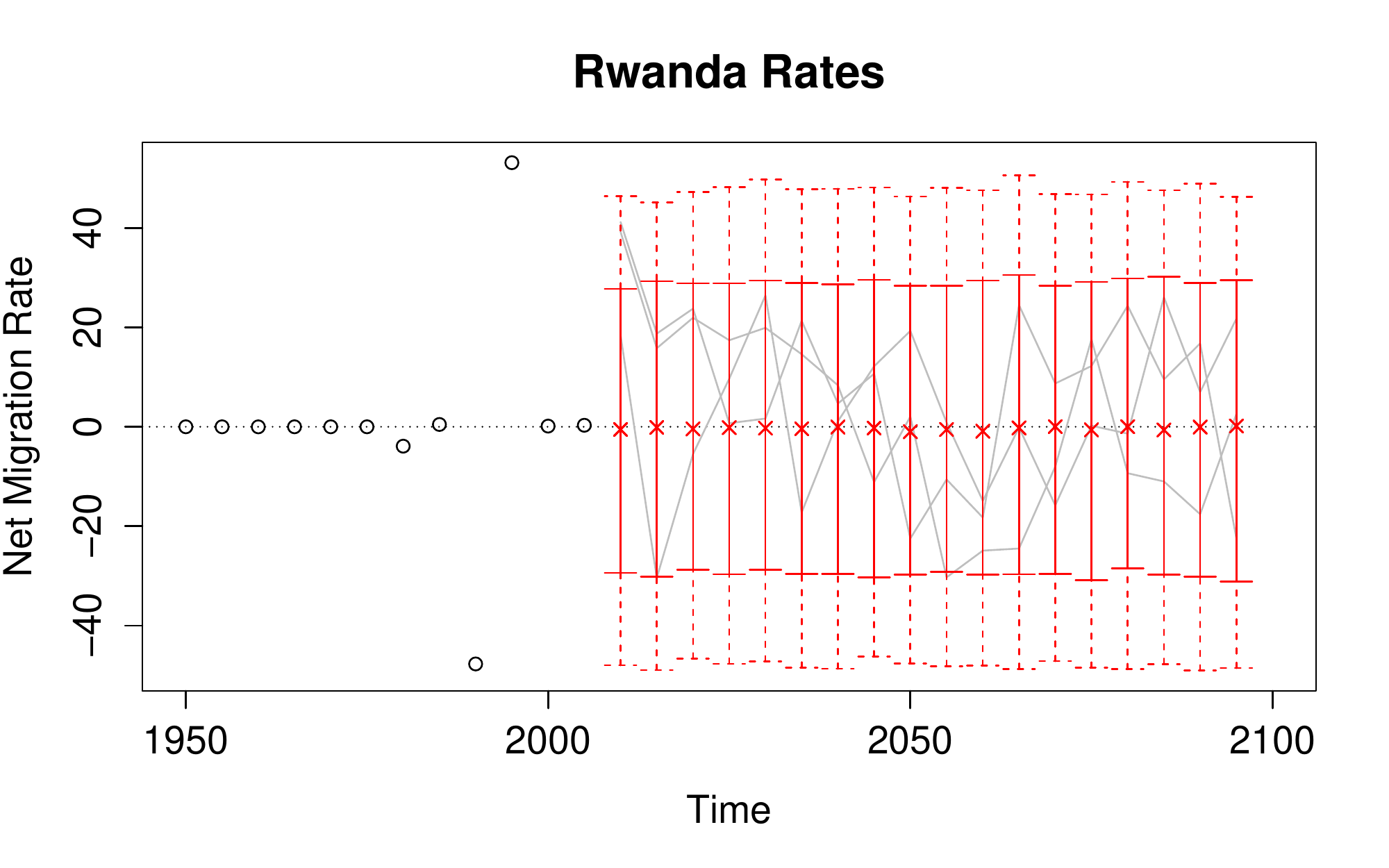}
\caption{Probabilistic Projections of Net International Migration Rates:
80\% and 95\% prediction intervals for Rwanda, with example trajectories included in gray.}
\label{fig:rwanda}
\end{figure}

Our model opts for wide prediction intervals in cases like Rwanda. A model which puts a heavy-tailed $t$ distribution on the $\varepsilon_{c,t}$'s rather than a normal distribution would produce narrower prediction intervals. However, we found that the normal model achieved better calibration. Section \ref{sec:discussion} contains a brief further discussion of a model with $t$-distributed errors.

\subsubsection{The least-developed countries}

The United Nations publishes a list of the least-developed countries, with countries classified as least-developed based on assessments of their economic vulnerability, human capital, and gross national income \cite{LDC2008}. A total of 46 countries in our data fall into the least-developed category. We now consider briefly the projections that our model makes for these least-developed countries in comparison to all other countries.

In the 2005-2010 time period, only 26\% of the least-developed countries were net receivers of migration, as compared to 43\% of all other countries. The least-developed countries had an average net migration rate of -0.97 per thousand, compared with an average of 2.64 per thousand in all other countries. However, our model projects that this gap in migration between currently least-developed and all other countries will narrow over time. Key findings are summarized in Table \ref{tab:projectedChange}. Over the coming decades, we project growth in net migration rates among the least developed countries and decline in net migration rate on average across all other countries.

\begin{table}[h!]
\centering
\caption{Mean projected change in migration rates (per thousand) among least-developed countries (LDC) versus all other countries (Other).}\label{tab:projectedChange}
\begin{tabular}{|l|c|c|}
\hline
 & LDC & Other\\
\hline
By 2020 & +0.02 & -1.49\\
By 2040 & +0.29 & -2.12\\
By 2060 & +0.34 & -2.29\\
\hline
\end{tabular}
\end{table}

\section{Discussion}\label{sec:discussion}
We have presented a method for projecting net migration rates. Our method is novel in that it provides probabilistic projections for all countries. Furthermore, it satisfies the requirement that simulated trajectories have zero global net migration for each sex and age group.

Additionally, we observe a paradoxical trend in the evolution of global migration rates. Although there is more migration than in the past as a proportion of the world population, countries' absolute migration rates have not been increasing on average. We resolve this paradox by noting the tendency of large countries to have small migration rates. Our method successfully reproduces this pattern, which seems desirable for migration projection methods in general.

Our model includes the assumption that the random error terms $\varepsilon_{c,t}$ are independent across countries and time. That assumption is mathematically convenient, but for many pairs of countries we expect to see non-zero correlations. For example, it is reasonable to expect that if Mexico undergoes particularly high net emigration during a quinquennium, then the United States will experience higher than usual net immigration during the same period. Thus we might expect to observe negative correlation between the random errors for Mexico and the United States. At the same time, it is not unreasonable to expect positive correlation between error terms in neighboring pairs of countries whose economic fortunes tend to move together. Such a pattern is observed, for example, among the Baltic states.
We attempted to find an optimal non-trivial covariance structure by constructing a variance-covariance matrix as a linear combination of matrices whose off-diagonal elements are pairwise, time-invariant covariates. However, this method offered no significant improvement over the assumption of independent residuals.

Migration rate data characteristically have outliers.
Wars and refugee movements, for example, produce migration rates which are on a much larger scale than are typical during times of stability. This suggests that a model with a long-tailed error distribution like a $t$ distribution might be more appropriate than a model with normal errors. However, in practice we found that models with normally distributed errors tended to outperform models with $t$ errors in out-of-sample predictive evaluation. Models with $t$ errors often produce 80\% and 95\% prediction intervals that are so tight that they do not come close to covering the range of observed historical migration rates. 
Statistically, the root of the problem is that in models with $t$ errors, large outliers often do not have a large effect on the inferred scale parameter. Although using $t$ errors often results in models with a high likelihood of the observed data, high likelihood does not necessarily correspond to good calibration of prediction intervals or qualitatively realistic migration rates. In our judgment, there is more value in forecasting distributions with reasonable prediction intervals than distributions which are likely to assign high probability density to future observations, and so we have used the normal model thoughout.

\paragraph{Acknowledgements}
This work was supported by the Eunice Kennedy Shriver National Institute of Child Health and Development through grants nos. R01 HD054511 and R01 HD070936, and by a Science Foundation Ireland E.~T.~S.~Walton visitor award, grant reference 11/W.1/I2079. The authors are grateful to Patrick Gerland and Joel Cohen for sharing data and helpful discussions.

\setlength\bibitemsep{1.5\itemsep}
\bibliographystyle{apacite}
\bibliography{bibliography}

\begin{thebibliography}{}

\bibitem [\protect \citeauthoryear {%
Abel%
}{%
Abel%
}{%
{\protect \APACyear {2010}}%
}]{%
abel2010}
\APACinsertmetastar {%
abel2010}%
\begin{APACrefauthors}%
Abel, G\BPBI J.%
\end{APACrefauthors}%
\unskip\
\newblock
\APACrefYearMonthDay{2010}{}{}.
\newblock
{\BBOQ}\APACrefatitle {Estimation of international migration flow tables in
  {Europe}} {Estimation of international migration flow tables in
  {Europe}}.{\BBCQ}
\newblock
\APACjournalVolNumPages{Journal of the Royal Statistical Society: Series A
  (Statistics in Society)}{173}{}{797--825}.
\PrintBackRefs{\CurrentBib}

\bibitem [\protect \citeauthoryear {%
Abel%
}{%
Abel%
}{%
{\protect \APACyear {2013}}%
}]{%
abel2013}
\APACinsertmetastar {%
abel2013}%
\begin{APACrefauthors}%
Abel, G\BPBI J.%
\end{APACrefauthors}%
\unskip\
\newblock
\APACrefYearMonthDay{2013}{}{}.
\newblock
{\BBOQ}\APACrefatitle {Estimating global migration flow tables using place of
  birth data} {Estimating global migration flow tables using place of birth
  data}.{\BBCQ}
\newblock
\APACjournalVolNumPages{Demographic Research}{28}{}{505--546}.
\PrintBackRefs{\CurrentBib}

\bibitem [\protect \citeauthoryear {%
Alkema%
\ \protect \BOthers {.}}{%
Alkema%
\ \protect \BOthers {.}}{%
{\protect \APACyear {2011}}%
}]{%
alkema2011}
\APACinsertmetastar {%
alkema2011}%
\begin{APACrefauthors}%
Alkema, L.%
, Raftery, A\BPBI E.%
, Gerland, P.%
, Clark, S\BPBI J.%
, Pelletier, F.%
, Buettner, T.%
\BCBL {}\ \BBA {} Heilig, G\BPBI K.%
\end{APACrefauthors}%
\unskip\
\newblock
\APACrefYearMonthDay{2011}{}{}.
\newblock
{\BBOQ}\APACrefatitle {Probabilistic projections of the total fertility rate
  for all countries} {Probabilistic projections of the total fertility rate for
  all countries}.{\BBCQ}
\newblock
\APACjournalVolNumPages{Demography}{48}{}{815--839}.
\PrintBackRefs{\CurrentBib}

\bibitem [\protect \citeauthoryear {%
Bijak%
\ \BBA {} Wi{\'s}niowski%
}{%
Bijak%
\ \BBA {} Wi{\'s}niowski%
}{%
{\protect \APACyear {2010}}%
}]{%
bijak2010}
\APACinsertmetastar {%
bijak2010}%
\begin{APACrefauthors}%
Bijak, J.%
\BCBT {}\ \BBA {} Wi{\'s}niowski, A.%
\end{APACrefauthors}%
\unskip\
\newblock
\APACrefYearMonthDay{2010}{}{}.
\newblock
{\BBOQ}\APACrefatitle {Bayesian forecasting of immigration to selected
  {European} countries by using expert knowledge} {Bayesian forecasting of
  immigration to selected {European} countries by using expert
  knowledge}.{\BBCQ}
\newblock
\APACjournalVolNumPages{Journal of the Royal Statistical Society: Series A
  (Statistics in Society)}{173}{}{775--796}.
\PrintBackRefs{\CurrentBib}

\bibitem [\protect \citeauthoryear {%
Br{\"u}cker%
\ \BBA {} Siliverstovs%
}{%
Br{\"u}cker%
\ \BBA {} Siliverstovs%
}{%
{\protect \APACyear {2006}}%
}]{%
brucker2006}
\APACinsertmetastar {%
brucker2006}%
\begin{APACrefauthors}%
Br{\"u}cker, H.%
\BCBT {}\ \BBA {} Siliverstovs, B.%
\end{APACrefauthors}%
\unskip\
\newblock
\APACrefYearMonthDay{2006}{}{}.
\newblock
{\BBOQ}\APACrefatitle {On the estimation and forecasting of international
  migration: {How} relevant is heterogeneity across countries?} {On the
  estimation and forecasting of international migration: {How} relevant is
  heterogeneity across countries?}{\BBCQ}
\newblock
\APACjournalVolNumPages{Empirical Economics}{31}{}{735--754}.
\PrintBackRefs{\CurrentBib}

\bibitem [\protect \citeauthoryear {%
Castles%
\ \BBA {} Miller%
}{%
Castles%
\ \BBA {} Miller%
}{%
{\protect \APACyear {2003}}%
}]{%
castles2003}
\APACinsertmetastar {%
castles2003}%
\begin{APACrefauthors}%
Castles, S.%
\BCBT {}\ \BBA {} Miller, M\BPBI J.%
\end{APACrefauthors}%
\unskip\
\newblock
\APACrefYear{2003}.
\newblock
\APACrefbtitle {The age of Migration: International Population Movements in the
  Modern World} {The age of migration: International population movements in
  the modern world}.
\newblock
\APACaddressPublisher{London}{Macmillan}.
\PrintBackRefs{\CurrentBib}

\bibitem [\protect \citeauthoryear {%
Cohen%
}{%
Cohen%
}{%
{\protect \APACyear {2012}}%
}]{%
cohen2012}
\APACinsertmetastar {%
cohen2012}%
\begin{APACrefauthors}%
Cohen, J\BPBI E.%
\end{APACrefauthors}%
\unskip\
\newblock
\APACrefYearMonthDay{2012}{}{}.
\newblock
{\BBOQ}\APACrefatitle {Projection of Net Migration Using a Gravity Model}
  {Projection of net migration using a gravity model}.{\BBCQ}
\newblock
\BIn{} \APACrefbtitle {Proc. {XXVII} {IUSSP} {International} {Population}
  {Conference}.} {Proc. {XXVII} {IUSSP} {International} {Population}
  {Conference}.}
\newblock
\APACrefnote{Retreived from
  \url{http://www.iussp.org/sites/default/files/event_call_for_papers/IUSSPsession020CohenProjectionNetMigrationGravityModelUNPopDiv2012corrected.pdf}}
\PrintBackRefs{\CurrentBib}

\bibitem [\protect \citeauthoryear {%
{Committee for Development Policy and United Nations Department of Economic and
  Social Affairs}%
}{%
{Committee for Development Policy and United Nations Department of Economic and
  Social Affairs}%
}{%
{\protect \APACyear {2008}}%
}]{%
LDC2008}
\APACinsertmetastar {%
LDC2008}%
\begin{APACrefauthors}%
{Committee for Development Policy and United Nations Department of Economic and
  Social Affairs}.%
\end{APACrefauthors}%
\unskip\
\newblock
\APACrefYear{2008}.
\newblock
\APACrefbtitle {Handbook on the Least Developed Country Category: Inclusion,
  Graduation, and Special Support Measures} {Handbook on the least developed
  country category: Inclusion, graduation, and special support measures}.
\PrintBackRefs{\CurrentBib}

\bibitem [\protect \citeauthoryear {%
Congdon%
}{%
Congdon%
}{%
{\protect \APACyear {2010}}%
}]{%
congdon2010}
\APACinsertmetastar {%
congdon2010}%
\begin{APACrefauthors}%
Congdon, P.%
\end{APACrefauthors}%
\unskip\
\newblock
\APACrefYearMonthDay{2010}{}{}.
\newblock
{\BBOQ}\APACrefatitle {Random-effects models for migration attractivity and
  retentivity: a {Bayesian} methodology} {Random-effects models for migration
  attractivity and retentivity: a {Bayesian} methodology}.{\BBCQ}
\newblock
\APACjournalVolNumPages{Journal of the Royal Statistical Society: Series A
  (Statistics in Society)}{173}{}{755--774}.
\PrintBackRefs{\CurrentBib}

\bibitem [\protect \citeauthoryear {%
Esipova%
, Ray%
\BCBL {}\ \BBA {} Publiese%
}{%
Esipova%
\ \protect \BOthers {.}}{%
{\protect \APACyear {2011}}%
}]{%
gallup2011}
\APACinsertmetastar {%
gallup2011}%
\begin{APACrefauthors}%
Esipova, N.%
, Ray, J.%
\BCBL {}\ \BBA {} Publiese, A.%
\end{APACrefauthors}%
\unskip\
\newblock
\APACrefYearMonthDay{2011}{}{}.
\newblock
{\BBOQ}\APACrefatitle {Gallup World Poll. {The} many faces of global migration}
  {Gallup world poll. {The} many faces of global migration}.{\BBCQ}
\newblock
\APACjournalVolNumPages{IOM Migration Research Series}{}{43}{}.
\PrintBackRefs{\CurrentBib}

\bibitem [\protect \citeauthoryear {%
Fertig%
\ \BBA {} Schmidt%
}{%
Fertig%
\ \BBA {} Schmidt%
}{%
{\protect \APACyear {2000}}%
}]{%
fertig2000}
\APACinsertmetastar {%
fertig2000}%
\begin{APACrefauthors}%
Fertig, M.%
\BCBT {}\ \BBA {} Schmidt, C\BPBI M.%
\end{APACrefauthors}%
\unskip\
\newblock
\APACrefYearMonthDay{2000}{}{}.
\newblock
\APACrefbtitle {Aggregate-level migration studies as a tool for forecasting
  future migration streams} {Aggregate-level migration studies as a tool for
  forecasting future migration streams}\ \APACbVolEdTR{}{\BTR{}}.
\newblock
\APACaddressInstitution{}{IZA Discussion paper series}.
\PrintBackRefs{\CurrentBib}

\bibitem [\protect \citeauthoryear {%
Hatton%
\ \BBA {} Williamson%
}{%
Hatton%
\ \BBA {} Williamson%
}{%
{\protect \APACyear {2002}}%
}]{%
hatton2002}
\APACinsertmetastar {%
hatton2002}%
\begin{APACrefauthors}%
Hatton, T\BPBI J.%
\BCBT {}\ \BBA {} Williamson, J\BPBI G.%
\end{APACrefauthors}%
\unskip\
\newblock
\APACrefYearMonthDay{2002}{}{}.
\newblock
\APACrefbtitle {What fundamentals drive world migration?} {What fundamentals
  drive world migration?}\ \APACbVolEdTR{}{\BTR{}}.
\newblock
\APACaddressInstitution{}{National Bureau of Economic Research}.
\PrintBackRefs{\CurrentBib}

\bibitem [\protect \citeauthoryear {%
Hatton%
\ \BBA {} Williamson%
}{%
Hatton%
\ \BBA {} Williamson%
}{%
{\protect \APACyear {2005}}%
}]{%
hatton2005}
\APACinsertmetastar {%
hatton2005}%
\begin{APACrefauthors}%
Hatton, T\BPBI J.%
\BCBT {}\ \BBA {} Williamson, J\BPBI G.%
\end{APACrefauthors}%
\unskip\
\newblock
\APACrefYear{2005}.
\newblock
\APACrefbtitle {Global migration and the world economy: Two centuries of policy
  and performance} {Global migration and the world economy: Two centuries of
  policy and performance}.
\newblock
\APACaddressPublisher{}{Cambridge Univ Press}.
\PrintBackRefs{\CurrentBib}

\bibitem [\protect \citeauthoryear {%
Hyndman%
\ \BBA {} Booth%
}{%
Hyndman%
\ \BBA {} Booth%
}{%
{\protect \APACyear {2008}}%
}]{%
hyndman2008}
\APACinsertmetastar {%
hyndman2008}%
\begin{APACrefauthors}%
Hyndman, R\BPBI J.%
\BCBT {}\ \BBA {} Booth, H.%
\end{APACrefauthors}%
\unskip\
\newblock
\APACrefYearMonthDay{2008}{}{}.
\newblock
{\BBOQ}\APACrefatitle {Stochastic population forecasts using functional data
  models for mortality, fertility and migration} {Stochastic population
  forecasts using functional data models for mortality, fertility and
  migration}.{\BBCQ}
\newblock
\APACjournalVolNumPages{International Journal of Forecasting}{24}{}{323--342}.
\PrintBackRefs{\CurrentBib}

\bibitem [\protect \citeauthoryear {%
Kim%
\ \BBA {} Cohen%
}{%
Kim%
\ \BBA {} Cohen%
}{%
{\protect \APACyear {2010}}%
}]{%
kim2010}
\APACinsertmetastar {%
kim2010}%
\begin{APACrefauthors}%
Kim, K.%
\BCBT {}\ \BBA {} Cohen, J\BPBI E.%
\end{APACrefauthors}%
\unskip\
\newblock
\APACrefYearMonthDay{2010}{}{}.
\newblock
{\BBOQ}\APACrefatitle {Determinants of International Migration Flows to and
  from Industrialized Countries: A Panel Data Approach Beyond Gravity}
  {Determinants of international migration flows to and from industrialized
  countries: A panel data approach beyond gravity}.{\BBCQ}
\newblock
\APACjournalVolNumPages{International Migration Review}{44}{}{899--932}.
\PrintBackRefs{\CurrentBib}

\bibitem [\protect \citeauthoryear {%
Lalic%
\ \BBA {} Raftery%
}{%
Lalic%
\ \BBA {} Raftery%
}{%
{\protect \APACyear {2012}}%
}]{%
LalicRaftery2012}
\APACinsertmetastar {%
LalicRaftery2012}%
\begin{APACrefauthors}%
Lalic, N.%
\BCBT {}\ \BBA {} Raftery, A\BPBI E.%
\end{APACrefauthors}%
\unskip\
\newblock
\APACrefYearMonthDay{2012}{}{}.
\newblock
\APACrefbtitle {Joint Probabilistic Projection of Female and Male Life
  Expectancy.} {Joint probabilistic projection of female and male life
  expectancy.}
\newblock
\APAChowpublished {Presented at the annual meeting of Population Association of
  America}.
\newblock
\APACrefnote{{http://paa2012.princeton.edu/abstracts/120140}}
\PrintBackRefs{\CurrentBib}

\bibitem [\protect \citeauthoryear {%
Lutz%
\ \BBA {} Goldstein%
}{%
Lutz%
\ \BBA {} Goldstein%
}{%
{\protect \APACyear {2004}}%
}]{%
lutz2004}
\APACinsertmetastar {%
lutz2004}%
\begin{APACrefauthors}%
Lutz, W.%
\BCBT {}\ \BBA {} Goldstein, J\BPBI R.%
\end{APACrefauthors}%
\unskip\
\newblock
\APACrefYearMonthDay{2004}{}{}.
\newblock
{\BBOQ}\APACrefatitle {Introduction: How to deal with uncertainty in population
  forecasting?} {Introduction: How to deal with uncertainty in population
  forecasting?}{\BBCQ}
\newblock
\APACjournalVolNumPages{International Statistical Review}{72}{}{1--4}.
\PrintBackRefs{\CurrentBib}

\bibitem [\protect \citeauthoryear {%
Massey%
\ \protect \BOthers {.}}{%
Massey%
\ \protect \BOthers {.}}{%
{\protect \APACyear {1993}}%
}]{%
massey1993}
\APACinsertmetastar {%
massey1993}%
\begin{APACrefauthors}%
Massey, D\BPBI S.%
, Arango, J.%
, Hugo, G.%
, Kouaouci, A.%
, Pellegrino, A.%
\BCBL {}\ \BBA {} Taylor, J\BPBI E.%
\end{APACrefauthors}%
\unskip\
\newblock
\APACrefYearMonthDay{1993}{}{}.
\newblock
{\BBOQ}\APACrefatitle {Theories of international migration: a review and
  appraisal} {Theories of international migration: a review and
  appraisal}.{\BBCQ}
\newblock
\APACjournalVolNumPages{Population and Development Review}{}{}{431--466}.
\PrintBackRefs{\CurrentBib}

\bibitem [\protect \citeauthoryear {%
Plummer%
}{%
Plummer%
}{%
{\protect \APACyear {2003}}%
}]{%
plummer2003}
\APACinsertmetastar {%
plummer2003}%
\begin{APACrefauthors}%
Plummer, M.%
\end{APACrefauthors}%
\unskip\
\newblock
\APACrefYearMonthDay{2003}{March}{}.
\newblock
{\BBOQ}\APACrefatitle {{JAGS}: A program for analysis of {Bayesian} graphical
  models using {Gibbs} sampling} {{JAGS}: A program for analysis of {Bayesian}
  graphical models using {Gibbs} sampling}.{\BBCQ}
\newblock
\BIn{} \APACrefbtitle {Proceedings of the 3rd International Workshop on
  Distributed Statistical Computing} {Proceedings of the 3rd international
  workshop on distributed statistical computing}\ (\BPGS\ 20--22).
\PrintBackRefs{\CurrentBib}

\bibitem [\protect \citeauthoryear {%
Raftery%
, Chunn%
, Gerland%
\BCBL {}\ \BBA {} {\v{S}}ev{\v{c}}{\'\i}kov{\'a}%
}{%
Raftery%
\ \protect \BOthers {.}}{%
{\protect \APACyear {2013}}%
}]{%
raftery2012life}
\APACinsertmetastar {%
raftery2012life}%
\begin{APACrefauthors}%
Raftery, A\BPBI E.%
, Chunn, J\BPBI L.%
, Gerland, P.%
\BCBL {}\ \BBA {} {\v{S}}ev{\v{c}}{\'\i}kov{\'a}, H.%
\end{APACrefauthors}%
\unskip\
\newblock
\APACrefYearMonthDay{2013}{}{}.
\newblock
{\BBOQ}\APACrefatitle {Bayesian probabilistic projections of life expectancy
  for all countries} {Bayesian probabilistic projections of life expectancy for
  all countries}.{\BBCQ}
\newblock
\APACjournalVolNumPages{Demography}{50}{}{777--801}.
\PrintBackRefs{\CurrentBib}

\bibitem [\protect \citeauthoryear {%
Raftery%
, Li%
, {\v{S}}ev{\v{c}}{\'\i}kov{\'a}%
, Gerland%
\BCBL {}\ \BBA {} Heilig%
}{%
Raftery%
\ \protect \BOthers {.}}{%
{\protect \APACyear {2012}}%
}]{%
raftery2012population}
\APACinsertmetastar {%
raftery2012population}%
\begin{APACrefauthors}%
Raftery, A\BPBI E.%
, Li, N.%
, {\v{S}}ev{\v{c}}{\'\i}kov{\'a}, H.%
, Gerland, P.%
\BCBL {}\ \BBA {} Heilig, G\BPBI K.%
\end{APACrefauthors}%
\unskip\
\newblock
\APACrefYearMonthDay{2012}{}{}.
\newblock
{\BBOQ}\APACrefatitle {Bayesian probabilistic population projections for all
  countries} {Bayesian probabilistic population projections for all
  countries}.{\BBCQ}
\newblock
\APACjournalVolNumPages{Proceedings of the National Academy of
  Sciences}{109}{}{13915--13921}.
\PrintBackRefs{\CurrentBib}

\bibitem [\protect \citeauthoryear {%
Raymer%
\ \BBA {} Rogers%
}{%
Raymer%
\ \BBA {} Rogers%
}{%
{\protect \APACyear {2007}}%
}]{%
raymer2007}
\APACinsertmetastar {%
raymer2007}%
\begin{APACrefauthors}%
Raymer, J.%
\BCBT {}\ \BBA {} Rogers, A.%
\end{APACrefauthors}%
\unskip\
\newblock
\APACrefYearMonthDay{2007}{}{}.
\newblock
{\BBOQ}\APACrefatitle {Using age and spatial flow structures in the indirect
  estimation of migration streams} {Using age and spatial flow structures in
  the indirect estimation of migration streams}.{\BBCQ}
\newblock
\APACjournalVolNumPages{Demography}{44}{}{199--223}.
\PrintBackRefs{\CurrentBib}

\bibitem [\protect \citeauthoryear {%
Richmond%
}{%
Richmond%
}{%
{\protect \APACyear {1988}}%
}]{%
richmond1988}
\APACinsertmetastar {%
richmond1988}%
\begin{APACrefauthors}%
Richmond, A\BPBI H.%
\end{APACrefauthors}%
\unskip\
\newblock
\APACrefYearMonthDay{1988}{}{}.
\newblock
{\BBOQ}\APACrefatitle {Sociological theories of international migration: the
  case of refugees} {Sociological theories of international migration: the case
  of refugees}.{\BBCQ}
\newblock
\APACjournalVolNumPages{Current Sociology}{36}{}{7--25}.
\PrintBackRefs{\CurrentBib}

\bibitem [\protect \citeauthoryear {%
Robertson%
}{%
Robertson%
}{%
{\protect \APACyear {1992}}%
}]{%
robertson1992}
\APACinsertmetastar {%
robertson1992}%
\begin{APACrefauthors}%
Robertson, R.%
\end{APACrefauthors}%
\unskip\
\newblock
\APACrefYear{1992}.
\newblock
\APACrefbtitle {Globalization: Social theory and global culture}
  {Globalization: Social theory and global culture}\ (\BVOL~16).
\newblock
\APACaddressPublisher{}{Sage}.
\PrintBackRefs{\CurrentBib}

\bibitem [\protect \citeauthoryear {%
Rogers%
\ \BBA {} Castro%
}{%
Rogers%
\ \BBA {} Castro%
}{%
{\protect \APACyear {1981}}%
}]{%
rogers1981}
\APACinsertmetastar {%
rogers1981}%
\begin{APACrefauthors}%
Rogers, A.%
\BCBT {}\ \BBA {} Castro, L\BPBI J.%
\end{APACrefauthors}%
\unskip\
\newblock
\APACrefYear{1981}.
\newblock
\APACrefbtitle {Model Migration Schedules} {Model migration schedules}.
\newblock
\APACaddressPublisher{Laxenburg, Austria}{International Institute for Applied
  Systems Analysis}.
\PrintBackRefs{\CurrentBib}

\bibitem [\protect \citeauthoryear {%
{ter Heide}%
}{%
{ter Heide}%
}{%
{\protect \APACyear {1963}}%
}]{%
heide1963}
\APACinsertmetastar {%
heide1963}%
\begin{APACrefauthors}%
{ter Heide}, H.%
\end{APACrefauthors}%
\unskip\
\newblock
\APACrefYearMonthDay{1963}{}{}.
\newblock
{\BBOQ}\APACrefatitle {Migration models and their significance for population
  forecasts} {Migration models and their significance for population
  forecasts}.{\BBCQ}
\newblock
\APACjournalVolNumPages{The Milbank Memorial Fund Quarterly}{41}{}{56--76}.
\PrintBackRefs{\CurrentBib}

\bibitem [\protect \citeauthoryear {%
{United Nations Population Division}%
}{%
{United Nations Population Division}%
}{%
{\protect \APACyear {2011}}%
}]{%
united2011}
\APACinsertmetastar {%
united2011}%
\begin{APACrefauthors}%
{United Nations Population Division}.%
\end{APACrefauthors}%
\unskip\
\newblock
\APACrefYear{2011}.
\newblock
\APACrefbtitle {World Population Prospects: The 2010 Revision} {World
  population prospects: The 2010 revision}.
\newblock
\APACaddressPublisher{}{United Nations}.
\PrintBackRefs{\CurrentBib}

\bibitem [\protect \citeauthoryear {%
{United Nations Population Division}%
}{%
{United Nations Population Division}%
}{%
{\protect \APACyear {2013}}%
}]{%
united2013}
\APACinsertmetastar {%
united2013}%
\begin{APACrefauthors}%
{United Nations Population Division}.%
\end{APACrefauthors}%
\unskip\
\newblock
\APACrefYear{2013}.
\newblock
\APACrefbtitle {World Population Prospects: The 2012 Revision} {World
  population prospects: The 2012 revision}.
\newblock
\APACaddressPublisher{}{United Nations}.
\PrintBackRefs{\CurrentBib}

\bibitem [\protect \citeauthoryear {%
Wheldon%
, Raftery%
, Clark%
\BCBL {}\ \BBA {} Gerland%
}{%
Wheldon%
\ \protect \BOthers {.}}{%
{\protect \APACyear {2013}}%
}]{%
Wheldon&2013}
\APACinsertmetastar {%
Wheldon&2013}%
\begin{APACrefauthors}%
Wheldon, M\BPBI C.%
, Raftery, A\BPBI E.%
, Clark, S\BPBI J.%
\BCBL {}\ \BBA {} Gerland, P.%
\end{APACrefauthors}%
\unskip\
\newblock
\APACrefYearMonthDay{2013}{}{}.
\newblock
{\BBOQ}\APACrefatitle {Estimating Demographic Parameters with Uncertainty from
  Fragmentary Data} {Estimating demographic parameters with uncertainty from
  fragmentary data}.{\BBCQ}
\newblock
\APACjournalVolNumPages{Journal of the American Statistical
  Association}{108}{}{96--110}.
\PrintBackRefs{\CurrentBib}

\bibitem [\protect \citeauthoryear {%
Wright%
}{%
Wright%
}{%
{\protect \APACyear {2010}}%
}]{%
wright2010}
\APACinsertmetastar {%
wright2010}%
\begin{APACrefauthors}%
Wright, E.%
\end{APACrefauthors}%
\unskip\
\newblock
\APACrefYearMonthDay{2010}{}{}.
\newblock
{\BBOQ}\APACrefatitle {2008-based national population projections for the
  {United Kingdom} and constituent countries} {2008-based national population
  projections for the {United Kingdom} and constituent countries}.{\BBCQ}
\newblock
\APACjournalVolNumPages{Population Trends}{139}{}{91--114}.
\PrintBackRefs{\CurrentBib}

\end{thebibliography}

\appendix
\section*{Appendix: Gravity Model Implementation}\label{appendix:cohen}

We implemented a version of \citeauthor{cohen2012}'s \citeyear{cohen2012} gravity model which projects net migration counts for five-year intervals starting at 2010 and ending at 2100. Projections are made for each country independently, with no redistribution step to ensure zero global net migration. For each country, projections are produced as follows: Let $L(t)$ be the population of country $c$ at time $t$ (in millions) and $M(t)$ be the population of the rest of the world at time $t$ (in millions). Then expected in-migration to country $c$ is given by $a \times L(t)^\alpha M(t)^\beta$, where $a$ is a country-specific proportionality constant and the exponents $\alpha$ and $\beta$ are constant across countries, with values estimated by \citeA{kim2010}. Similary, expected out-migration from country $c$ has the form $b\times L(t)^\gamma M(t)^\delta$, where $b$ is to be estimated and $\gamma$ and $\delta$ come from \citeA{kim2010}. 

The constants of proportionality $a$ and $b$ for each country are chosen to minimize the sum of squared deviations between estimates of net migration from the gravity model and WPP estimates of net migration \cite{united2011} given in units of millions of net annual migrants. We used the values $\alpha=0.728$, $\beta=0.602$, $\gamma=0.373$, and $\delta=0.948$, reported by \citeA{cohen2012}. For each country, having estimated $a$ and $b$, net migration projections are then given by $a \times L(t)^\alpha M(t)^\beta - b\times L(t)^\gamma M(t)^\delta$, where $L(t)$ and $M(t)$ are now projected populations also taken from WPP's 2010 revision \cite{united2011}.

\begin{figure}
\centering
\includegraphics[width=0.6\textwidth]{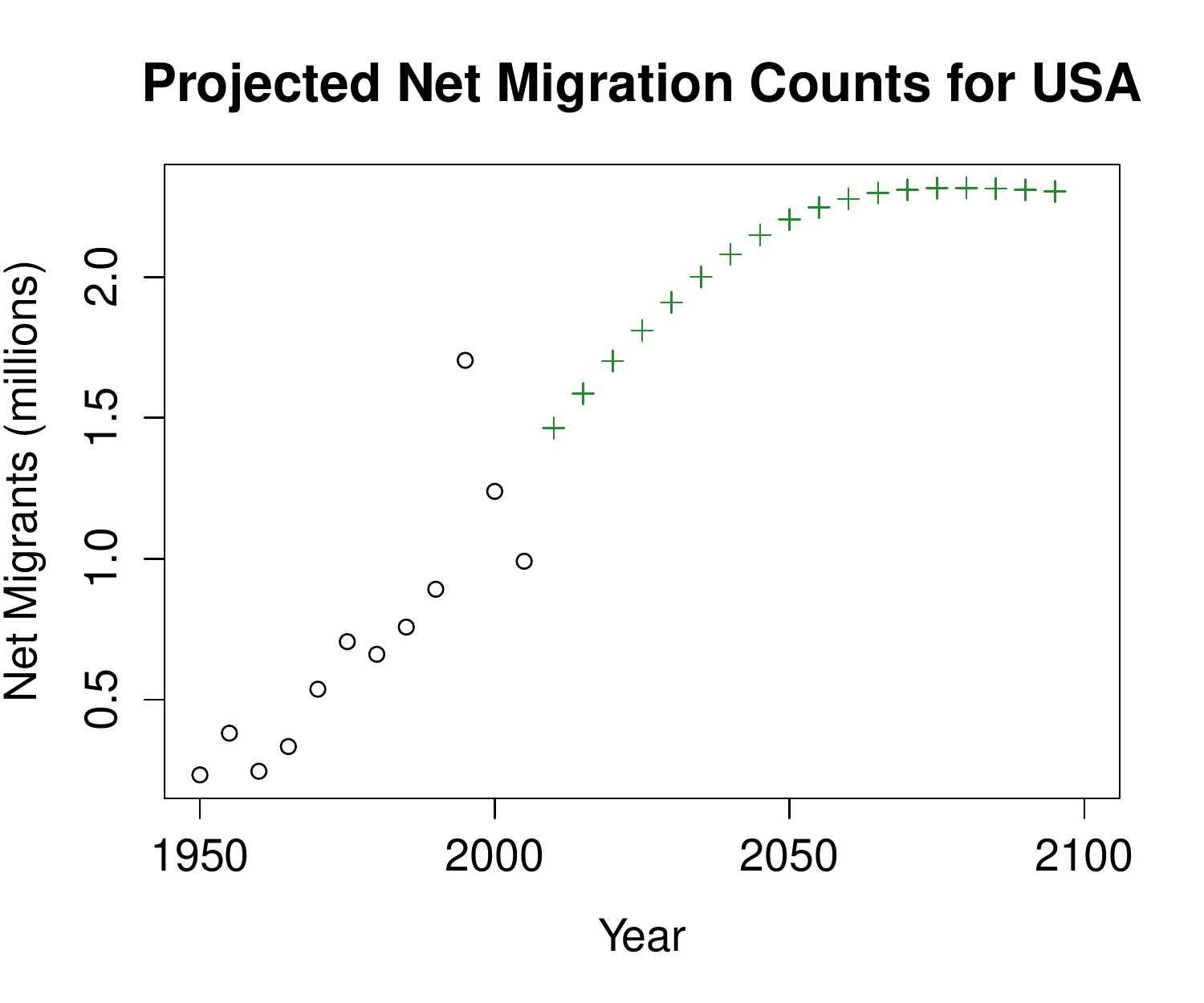}
\caption{Gravity model based projections of net international migration counts for the USA.}
\label{fig:gravity}
\end{figure}

Our implementation appears to reproduce the results in \citeA{cohen2012}. \citeauthor{cohen2012} reports the values of the proportionality constants, $a$ and $b$, obtained for the United States, and provides a plot of the projections from his implementation of the gravity model. Using these, we are able to confirm that our results agree with those from \citeauthor{cohen2012}'s implementation. \citeauthor{cohen2012} reports $a=3.43 \times 10^{-4}$ and $b=-8.28 \times 10^{-4}$. We find very similar values of $a=3.42 \times 10^{-4}$ and $b=-8.33 \times 10^{-4}$. The slight discrepancies may come from having used only three decimal places of the values for $\alpha$, $\beta$, $\gamma$, and $\delta$ in our implementation. Furthermore, Figure \ref{fig:gravity} shows the projected net migration counts for the United States using our implementation of the gravity model. Our projections appear to be essentially the same as the gravity model projections plotted in Figure 1(b) of \citeA{cohen2012}.

\end{document}